\documentclass[
aip,
amsmath,amssymb,
reprint,
]{revtex4-1}

\usepackage{mathptmx}
\usepackage[scaled=0.9]{helvet}
\usepackage{xcolor}
\usepackage{graphicx}
\usepackage{dcolumn}
\usepackage[flushleft]{threeparttable}
\usepackage{multirow}
\usepackage{array}

\renewcommand{\=}{\mbox{\,=\,}}
\newcommand{\e}{\mathrm{e}}
\renewcommand{\i}{\mathrm{i}}
\renewcommand{\d}{\mathrm{d}}

\begin{document}

\author{Sajal Kumar Giri}
\affiliation{Department of Chemistry, Northwestern University, 2145 Sheridan Road, Evanston, Illinois 60208, United States}
\author{George C. Schatz}
\email{g-schatz@northwestern.edu}
\affiliation{Department of Chemistry, Northwestern University, 2145 Sheridan Road, Evanston, Illinois 60208, United States}
\title{Laser Pulse Induced Second- and Third-Harmonic Generation of Gold Nanorods with Real-Time Time-Dependent Density Functional Tight Binding (RT-TDDFTB) Method}

\begin{abstract} 
In this study, we investigate second- and third-harmonic generation processes in Au nanorod systems using the real-time time-dependent density functional tight binding (RT-TDDFTB) method. 
Our study focuses on computation of nonlinear signals based on the time dependent dipole response induced by linearly polarized laser pulses interacting with nanoparticles. 
We systematically explore the influence of various laser parameters, including pump intensity, duration, frequency, and polarization directions, on the harmonic generation. 
We demonstrate all the results using Au nanorod dimer systems arranged in end-to-end configurations, and disrupting the spatial symmetry of regular single nanorod systems crucial for second harmonic generation processes.
Furthermore, we study the impact of nanorod lengths, which lead to variable plasmon energies, on the harmonic generation, and estimates of polarizabilities and hyper-polarizabilities are provided.
\end{abstract}
\maketitle

\section{Introduction}
\noindent The study of nonlinear optical processes at the nanoscale has gained substantial interest in recent years due to its importance in various fields, including microscopy,\cite{Nonlinear-Plasmon-Microscopy-2007-Sun, Nonlinear-Plasmon-Microscopy-2010-Fraser, Nonlinear-Plasmon-Microscopy-2019-Zheng} optoelectronics,\cite{Nonlinear-Plasmon-Optoelectronics-2012-Loh, Nonlinear-Plasmon-Optoelectronics-2022-Yao} and biomedical applications.\cite{Nonlinear-Plasmon-Biology-2003-Loew, SHG-Bio-2012-Pantazis, Nonlinear-Plasmon-Biology-2017-Chu}
Among these phenomena, harmonic generation (HG) has emerged as an important process for manipulating light at the nanoscale.\cite{Hyper-Rayleigh-2002-Schtaz, Nonlinear-Plasmon-2002-Schatz, HG-Plasmon-2008-Kim, Nonlinear-Plasmon-2012-Zayats, Nonlinear-Plasmon-2020-Celebrano, Nonlinear-Plasmon-2023-Giessen} 
This process involves the nonlinear interaction between intense laser fields and matter, where the induced polarization scales non-linearly with the interacting field amplitude, resulting in the emission of harmonics at integer multiples of the incident laser frequency.\cite{HG-1961-Weinreich, HG-1962-Pershan} \\

Most of the past studies on HG have involved atomic\cite{HG-Noble-1992-Sakai, HG-Noble-1993-Balcou, HG-Noble-1993-Gordon} or molecular\cite{HG-Mol-1994-Chaker, HG-Mol-1996-Wahlstrom} systems.
However, the generated signals are usually extremely weak, requiring very intense incident pulses. 
This is because the HG process involves multiple photons, and there is a low conversion efficiency associated with the weak light-matter coupling and a low density of states of the medium.
There is an ongoing effort to identify ideal target systems for efficient HG processes. 
Recent experimental successes in HG on liquids,\cite{HG-Liq-2018-Worner} solids,\cite{HG-Solids-2019-Reis} and nanostructures\cite{HG-Nano-2020-Kivshar} have significantly enhanced the conversion efficiency beyond the limits of gas target systems. \\

Second and third harmonic generation processes are of particular interest due to their emerging potential applications that cannot be achieved using conventional linear absorption.\cite{Plasmon-SHG-2015-Martin, Plasmon-SHG-2024-Wang}
Second harmonic generation (SHG) is a nonlinear optical process that involves the conversion of the fundamental frequency into second harmonics at doubled the frequency. 
SHG involving nanoparticles in solution involves the coherent excitation of each particle.
However, the contributions from the ensemble of particles that are irradiated involve an incoherent sum, leading to a process that is sometimes referred to as Hyper-Rayleigh scattering, but we will use the term SHG throughout this paper.
Under the dipole approximation, which applies when the dimension of a particle is much smaller than the wavelength of the incident radiation, centrosymmetric particles are inactive in SHG experiments,\cite{HG-1961-Weinreich, HG-1962-Pershan} as long as the particles remain much smaller than the wavelength of light.
SHG emission from small nanoparticles is, therefore, available only for structures with asymmetric geometries, for example, a small metallic particle with the shape deviating from a perfect sphere.\cite{SHG-Sphere-2005-Brevet}
Similarly, third harmonic generation (THG) involves the tripling of the incident frequency, and is allowed for spherical particles.\cite{HG-1961-Weinreich, HG-1962-Pershan} \\

Noble metal nanoparticles, in particular, exhibit enhanced optical absorption due to collective excitation of conduction electrons when excited with a characteristic frequency of the particles, a phenomenon known as localized surface plasmon resonance (LSPR).\cite{LSPR-2007-VanDuyne, LSPR-2003-Schatz}  
These plasmonic resonances can significantly enhance the local electromagnetic field, thereby intensifying the nonlinear optical processes involved in HG. 
SHG\cite{Nonlinear-Plasmon-SHG-1981-Shen, Plasmon-SHG-THG-2009-Leitenstorfer, Plasmon-SHG-2010-Brevet, Plasmon-SHG-2015-Arbouet, Plasmon-SHG-2015-Giessen, Plasmon-SHG-2020-Ciraci, HG-2021-Giessen} and THG\cite{THG-Plasmon-2005-Orrit, Plasmon-SHG-THG-2009-Leitenstorfer, Plasmon-THG-2017-Maier, THG-2019-Sukharev} can be enhanced by several orders of magnitude due to the surface plasmon effects.
Here, plasmonic excitation provides both stronger coupling and high state densities for the conversion of the incident photons into higher-energy harmonics. 
The unique tunability of nanoparticle plasmons allows for precise control over the harmonic generation process, offering opportunities for tailoring the spectral and spatial properties of generated harmonics. 
Moreover, plasmonic nanoparticles provide a significant advantage compared to harmonic generation from bulk solids in achieving phase matching conditions between the incident and generated harmonics, which is essential for an efficient harmonic emission. \\

Previous theoretical treatments of the optical properties of nanoparticles have employed several methods, including simulations based on Maxwell's equations,\cite{SHG-1994-Schatz} mixed quantum-classical treatments,\cite{Quantum-Classical-2010-Schatz, Quantum-Classical-2012-Schatz, Quantum-Classical-2012-Neuhauser} time-dependent density functional theory (TD-DFT)\cite{TDDFT-2006-Schatz-2, TDDFT-2006-Schatz, Quantum-2012-Borisov, TDDFT-2017-Borisov, TDDFT-HG-2020-Borisov} and semi-empirical methods\cite{INDO-2016-Schatz, INDO-2017-Schatz}. 
However, only a few of these methods have been used to determine HG properties for metallic nanoparticles.
Finite-difference time-domain (FDTD) simulations offer a powerful tool for modeling the interaction of intense laser fields with nanoparticles, providing insights into the spatiotemporal evolution of electromagnetic fields and the resulting harmonic generation.\cite{FDTD-2009-Neuhauser, FDTD-2011-Liu, FDTD-2016-Sukharev, FDTD-2017-Sukharev, FDTD-2021-Sadrnezhaad, FDTD-2024-Sukharev} 
As an alternative to electrodynamics, TD-DFT calculations provide a more rigorous picture of the electronic excitation and transitions in nanoparticles, elucidating the role of electronic structure in HG processes.
However, computational efficiency is a major concern while using TD-DFT for nanoparticles. 
Alternatively, semi-empirical methods such as INDO/S provide computationally efficient approaches for investigating nonlinear optical processes in complex nanoparticle systems, however, computational constraints have still limited the exploration of large-scale applications. \\

The density functional tight binding (DFTB) method is a semi-empirical approach that shares similarities with the earlier INDO-based methods but is formulated within the framework of density functional theory (DFT), providing improved accuracy.
Indeed, this method often generates results comparable to DFT but with significantly reduced computational costs, enabling efficient simulation of ground and excited state properties for larger complexes. 
However, the accuracy of predictions using this method relies on the quality of parameters used to represent the Hamiltonian ($H$) and overlap ($\mathcal{O}$) matrices.
Recent papers provide comprehensive details of the DFTB method.\cite{DFTB-2020-Aradi,RTDFTB-2020-Sanchez} 
Previous studies have demonstrated the success of DFTB for predicting the structural, thermodynamic, and optical properties of various systems, including bio-molecules,\cite{DFTB-Bio-2014-Marcus,DFTB-Bio-2019-Wong} metals,\cite{DFTB-Metal-2019-Bossche,DFTB-Plasmon-2020-Visscher,DFTB-Plasmon-2022-Sala} semiconductors,\cite{DFTB-SemiConductor-2022-Heine} etc. \\

The real-time time-dependent density functional tight binding (RT-TDDFTB) method extends the ground state DFTB approach into the time domain, explicitly incorporating interactions with external driving fields.
This framework naturally accommodates light-induced processes, making it particularly suited for studying the dynamic behavior of materials under optical excitation.\cite{RTDFTB-2020-Sanchez,RT-TDDFTB-2023-Wong}
As a result, RT-TDDFTB has been used in simulating the optical responses of nanostructures, such as plasmon resonances,\cite{RTDFTB-2016-Sanchez,RT-TDDFTB-Hot-2019-Sanchez} hot-carrier generation,\cite{RT-TDDFTB-Hot-2019-Sanchez,DFTB-PDC-2022-Zhang,RT-TDDFTB-Hot-2022-Sanchez} photocatalysis,\cite{DFTB-PDC-2023-Schatz,DFTB-PDC-2023-Zhang,DFTB-SERS-2024-Schatz} and excitation energy transfer\cite{RT-TDDFTB-2017-Wong,RT-DFTB-2018-Wong,RT-TDDFTB-2023-Aikens}.  \\

In this work, we explore SHG and THG processes in end-to-end Au nanorod dimer systems induced by linearly polarized laser pulses using the real-time TD-DFTB method through the computation of time-dependent induced dipoles. 
This paper includes two major objectives: (a) to determine the feasibility of studying nonlinear optical processes for plasmonic nanoparticles by simulating real-time dynamics at the atomistic level and (b) to assess the influence of laser pulse parameters on these processes.
We present a comprehensive analysis of the effect of various laser parameters, including pump intensity, frequency, duration, and laser polarization direction, on the nonlinear optical phenomena. We also consider the influence of plasmon dephasing and damping on the results.
Additionally, we investigate the impact of nanoparticle size, determining how the length of the nanorods affects the HG efficiency, and we report the polarizabilities and first and second hyper-polarizabilities for dimer systems calculated from the induced dipoles, including comparisons with the experiment.  
The results demonstrate that the real-time TD-DFTB provides a powerful tool for determining HG properties in plasmonic nanoparticles.\\

Throughout this paper, atomic units are utilized unless otherwise specified.

\section{Theory and Computation}
\noindent All computations in this study have been conducted using the DFTB+ code that works based on the DFTB method.
We use auorg/auorg-1-1 Slater-Koster parameters for Au systems downloaded from the dftb.org site. 
This method provides a balance between accuracy and computational cost, making it particularly suitable for large-scale simulations.
The accuracy of the DFTB calculations is typically comparable to that of the more rigorous DFT method while offering substantially reduced computational costs.
However, it is important to note that the accuracy of these calculations can be influenced by various factors, including the empirical parameters used to describe the electronic structure.
In addition to investigating ground state properties, we used this method to study excited state properties that is an extension of the method to the time domain. 
Furthermore, this method allows us to incorporate external electromagnetic fields explicitly to drive the dynamics through the light-matter interactions. \\

In real-time dynamics simulations within the DFTB framework, the evolution of the single particle electronic density matrix is governed by the Liouville von-Neumann equation.\cite{RTDFTB-2020-Sanchez} 
The density matrix is propagated in time using
\begin{equation}
    \dot{\rho}(t)\=-{\rm i}(\mathcal{O}^{-1}H\rho-\rho H\mathcal{O}^{-1}),
    \label{lvn_eq}
\end{equation}
starting from an initial state defined by Hamiltonian ($H_0$), overlap ($\mathcal{O}_0$), and density ($\rho_0$) matrices.
Here $\mathcal{O}^{-1}$ represents the inverse of $\mathcal{O}$.
The above equation describes the propagation of the electronic density matrix over time where the interaction with an external field initiates transitions between electronic states. 
The Hamiltonian includes a light-matter coupling term through the dipole approximation $V_A(t)=-\boldsymbol{\mu}\cdot\mathbf{E}(t)$, where $\mathbf{E}(t)$ is the external field and $\boldsymbol{\mu}=\sum_A\Delta q_A \mathbf{R}_A(t)$ is the dipole moment associated with atomic coordinate $\mathbf{R}_A$ and charge $\Delta q_A$ for atom A at time $t$.
The external field takes the following form:
\begin{equation}
    E_a(t)\=\hat{e}_aE_p{\rm exp}(-(t-t_c)^2/T_p^2)\sin(\omega_pt),
    \label{pulse_eq}
\end{equation}
with peak amplitude $E_p$, center of the pulse $t_c$, pulse duration $T_p$, frequency $\omega_p$, and polarization direction $\hat{e}_a$.
By solving Eq.\ref{lvn_eq} numerically, it becomes possible to track the dynamic behavior of the system as it evolves in real time. 
Although the light-matter interaction may trigger nuclear dynamics, our focus lies on the electronic dynamics for HG processes, hence we constrain nuclear motion in our simulation. 
This is a legitimate approximation as nuclei initially exhibit slow movement, and we only need to simulate dynamics up to 200 fs for the properties we are calculating.\cite{RTDFTB-2016-Sanchez, RTDFTB-2016-Sanchez-2, RTDFTB-2020-Sanchez} 
Nuclear dynamics becomes more relevant for materials over longer timescales. \\

To elucidate the process of HG in nanoparticles, we excite them with an intense laser pulse of the form given in Eq.\ref{pulse_eq}. 
The induced dipole $\mu_a(t)={\rm Tr}[(\hat{e}_a\cdot\boldsymbol{\mu})\rho(t)]$ oscillates in time coherently for an infinitely long time in the absence of dissipation, including all frequency modes that are activated through the interactions, including both fundamental and generated harmonics. 
The induced dipole can be expressed as\cite{RT-HyperPol-2013-Li} 
\begin{align} \nonumber
    \mu_a(t)&\=\mu_a^0+\sum_b\int_{-\infty}^\infty\d t_1 \alpha_{ab}(t-t_1)E_b(t_1) \\ \nonumber
    &+\frac{1}{2!}\sum_{bc}\int\int_{-\infty}^\infty\d t_1\d t_2 \beta_{abc}(t-t_1, t-t_2)E_b(t_1)E_c(t_2) \\ \nonumber
    &+\frac{1}{3!}\sum_{bcd}\int\int\int_{-\infty}^\infty\d t_1\d t_2\d t_3 \gamma_{abcd}(t-t_1, t-t_2, t-t_3) \\ 
    &\times E_b(t_1)E_c(t_2)E_d(t_3) +\ldots,
\label{dipole_eq}
\end{align}
where $a,\,b,\,c$ and $d$ denote Cartesian coordinates. 
$\alpha$, $\beta$ and $\gamma$ are hyper-polarizability tensors representing the first, second, and third order responses, as determined by the polarizability and the first and second hyper-polarizabilities, respectively.
In the above equation, $\mu_a^0$ is the permanent dipole moment. \\

To provide a realistic description of the excited state dynamics that includes dephasing and relaxation due to electron-electron and electron-nuclear interactions, we introduce empirical damping to the dipole amplitude, effectively limiting the oscillation to a finite time.\cite{RTDFTB-2016-Sanchez, RT-TDDFT-2020-Kuisma} 
This damping function is chosen to be an exponential function of the form given in Eq.\ref{hg_damp_eq}, where $\tau$ controls the dephasing/damping rate. 
Throughout the paper we use $\tau=20$ fs. 
This value is related to the Drude width and often this is considered to be  $\sim\,0.1$ eV corresponding to $\tau=10$ fs.\cite{Maxwell-1987-Schatz}
This has been used in the past in electrodynamics calculations to predict extinction spectra for spherical gold nanoparticles for wavelengths close to 520 nm, where the measured width of the plasmon is around 0.2 eV (where deviations between the two widths are due to effects not contained in the Drude model as well as uncertainties in determining values of the widths from the measured lineshape due to a nonresonant background).
For rod-shaped particles where the plasmon is shifted to a longer wavelength (in the NIR), the Drude width based on fitting measured dielectric functions is usually given a smaller value, such as 0.02 eV.\cite{LS-2009-Ford}
However, for rods that are large (10s of nm in diameter), the plasmon width based on electrodynamics calculations is found to slowly decrease from 0.2 to 0.1 eV in going from 520 nm to over 1000 nm. 
This arises because radiative damping contributions broaden the plasmon beyond what it would be due to intrinsic damping, and there are also important frequency dispersion effects that control the linewidth.\cite{LS-2010-Schatz, LS-2011-Schatz}
In the present application we are interested in using our electronic structure calculations for small rods to describe properties that can be connected to the behavior of large rods, so we use $\tau=20$ fs as a value that is more realistic for the plasmon resonances we consider, but without putting in an explicit dependence of $\tau$ on plasmon wavelength.
The influence of other values of $\tau$ will be presented later when we consider polarizability results.  \\

The dipole oscillations are primarily dominated by the fundamental frequency of the laser excitation.
However, all the signals, including the generated harmonics, are obtained from the Fourier transform of the induced dipole $\mu(t)$:\cite{HG-Theory-2021-Luppi}
\begin{align}
    \label{hg_eq}
    S(\omega)&\=\Big{|}\sum_a\int_{0}^{t_f}\d t\,g(t)[\mu_a(t)-\mu_a^0]\e^{-\i\omega t}\Big{|}^2, \\
      g(t)&\=\begin{cases}
        \e^{-(t-t_s)/\tau}, & \text{for $t\,\geq\,t_s$},\\
        1, & \text{otherwise},
  \end{cases}
    \label{hg_damp_eq}
\end{align}
where the pulse amplitude decays to zero at time $t_s$, and $t_f$ is the final propagation time.
Throughout our analysis, we ensure that the final time $t_f$ ($t_f\approx\,200$ fs) is selected such that the dipole amplitude decays to zero for the specified damping parameters.\\

The linear absorption spectrum is computed from the induced real-time dipole. 
To access a broad energy range, nanoparticles are subjected to interact with a Dirac delta perturbation in the form of $E_a^\delta(t)\=E_{p,a}^{\delta}\delta(t-t')$, where $E_{p,a}^\delta$ is the field amplitude component for the polarization direction $\hat{e}_a$, instead of interacting with a laser pulse as used for harmonic signals.  
For harmonic generation, the purpose of using a pulsed signal is to excite nanoparticles within a well-defined energy range contained within the pulse. 
This allows for the separation of each harmonic from one another, with harmonics spaced by the driving frequency. This type of calculation can be used to calculate hyper-polarizabilities as described later.
From the induced dipole, the polarizability ($\alpha_{ab}$) component and absorption cross-section ($\sigma$) are computed as
\begin{align}
    \label{alpha_eq} 
    \alpha_{ab}(\omega)&\=\frac{\mu_{a}^\delta(\omega)}{E_{p,b}^\delta},\\
    \sigma(\omega)&\=\frac{4\pi\omega}{3c}\Im[\alpha_{xx}+\alpha_{yy}+\alpha_{zz}],
    \label{abs_eq} 
\end{align}
where $\Im[\cdot]$ denotes the imaginary part of the argument.
Here $\mu_a^\delta(\omega)$ is the Fourier transform of the time-dependent dipole $\mu_a^\delta(t)$ which includes damping over time,
given by $\mu_a^\delta(\omega)\=\int_{-\infty}^\infty\d\,t\,[\mu_a^\delta(t)-\mu_a^0]\times\e^{-t/\tau}\e^{\i\omega\,t}$. \\

Polarizability determines absorption, while the first and second hyper-polarizabilities measure the efficiency of SHG and THG for a given particle. 
All polarizabilities in this paper are calculated according to Eq.\ref{alpha_eq}, unless explicitly specified otherwise. 
However for hyper-polarizabilities, we use a modified approach based on the method developed by Li and co-workers \cite{RT-HyperPol-2013-Li}, which involves extracting polarizabilities and hyper-polarizabilities from the real-time dipole. 
This method allows us to obtain dipole components of different orders, followed by deriving polarizability and hyper-polarizability components through fitting with analytical functions. 
The Li method works well for nonresonant excitation of small molecules, but for the present case where we have resonant excitation of a plasmonic particle with a large density of states, we find it necessary to modify the method to employ a laser pulse of the form given in Eq.\ref{pulse_eq}.

\section{Results and Discussion}
\subsection*{A. Systems and Absorption Spectra} 
\subsubsection*{1. Systems}
\begin{figure}[h]
    \centering
    \includegraphics[width = 0.4 \textwidth]{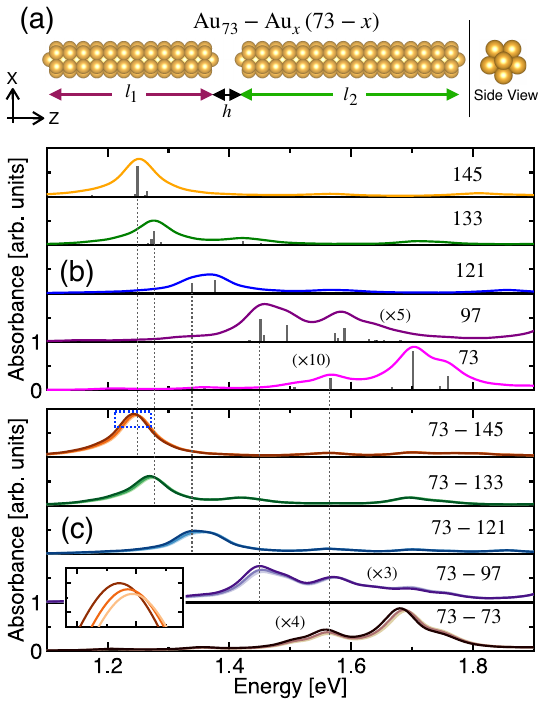}
    \caption{(a) Nanorod dimer systems used in this study. 
    The two nanoparticles are characterized by lengths $l_1$ and $l_2$ respectively, with an inter-spacing of $h$ between them.
    Absorption spectra for (b) single nanorod and (c) dimer systems. 
    For each system in (c), the inter-particle spacing increases from 0.6 to 1 nm as shown by the color change from deep to light.
    Spectra are obtained from real-time dipole calculations through Eq.\ref{abs_eq}.
    In (b), vertical solid lines denote oscillator strengths connecting the ground and excited states through one-photon processes.
    Spectra are amplified by factors of 10(4) and 5(3) for the 73(73-73) and 97(73-97) systems, respectively, for comparison. 
    The inset in (c) magnifies the peaks for different $h$ values for the 73-145 system within the region indicated by the blue dashed box.
    Here, the peak intensity of delta kick is $I_p=1.3\times10^{7}$ W/cm$^2$, for calculating the real-time absorption spectra. 
    } 
   \label{sys_fig}
\end{figure}

\noindent Figure \ref{sys_fig}(a) illustrates the Au nanorod systems used in this study. 
The symmetry of nanoparticle and polarization direction play important roles for SHG.\cite{HG-1962-Pershan} 
For a single nanorod, there is symmetry with respect to inversion along the rod axis, but not in the transverse direction (due to the five Au atoms that are around the central axis of the particle, Fig.\ref{sys_fig}(a) side view).
As a result, the rod should be inactive for SHG when excited along the rod axis for a linearly polarized laser pulse along that axis, as will be demonstrated later.  \\

To generate asymmetric arrangements of Au atoms along the longitudinal direction, we investigate end-to-end dimer nanorod systems of different lengths as shown in Figure \ref{sys_fig}.
Previous experimental studies of much larger rods have reported the synthesis of end-to-end nanorod systems\cite{Rod-2003-Murphy, Rod-2022-Orrit} and investigated nonlinear optical processes using such configurations\cite{Nonlinear-Gold-2009-Leitenstorfer, Nonlinear-Gold-Nanowire-2012-Novotny}.
In this study, the dimer geometries are generated using monomer units that were optimized with a maximum force tolerance of  10$^{-3}$ eV/$\AA$.
To generate a closed shell configuration, we use a -1 charge for each nanorod system, resulting in dimer configurations with a total charge of -2. 
We vary the length of one nanorod system while keeping the length of the other particle unchanged to manipulate the asymmetry i.e. the length ratio between the two particles.
Five different length combinations are studied, denoted as Au$_{73}$-Au$_x$, with $x$ taking the values 73, 97, 121, 133, and 145.
Therefore, we have a symmetric configuration only for $x=73$ and asymmetric configurations for the rest of the values of $x$.
Here, we use the monomer particle involving 73 atoms (with a length of 3.2 nm) as a reference; however, our findings remain consistent with other particle sizes as references.
Different inter nanoparticle spacings ($h$=0.6, 0.8, and 1 nm) are used to modulate interactions between particles. 
Note that the smallest value chosen is large enough to avoid wavefunction overlap effects that would change the optical response in significant ways.
It is worth mentioning that while other dimer geometries (like T-shaped\cite{SHG-T-2007-Kauranen, SHG-L-T-2015-Kauranen}, L-shaped\cite{SHG-L-2007-Turunen, SHG-L-T-2015-Kauranen}) could alternatively provide spatial asymmetry, for computational simplicity, we restrict this study to the end-to-end dimer configurations.  \\

\subsubsection*{2. Absorption Spectra}
\noindent Absorption spectra of the nanorod systems are shown in Figure \ref{sys_fig}(b, c), as obtained from the real-time calculations through computation of the time-dependent induced dipoles.
Additionally, we computed absorption spectra for some systems in the frequency domain using linear response TD-DFTB.
The time and frequency domain results agree with each other, providing a validation of the real-time calculations in this study.
However, the frequency domain calculations are much more computationally expensive compared to real-time calculations, particularly when a wide energy window and a substantial number of excited states are needed to cover the energy range.
Given this, we performed frequency domain computations only for single nanorod geometries (elements of the nanorod dimers) and compared the results with the spectra obtained from real-time calculations, as shown in Fig.\ref{sys_fig}(b).  \\

The dashed vertical lines in Figure \ref{sys_fig}(b) indicate the maximum absorption peaks, which represent excitation of the longitudinal plasmon mode within each particle.
The involved excited states exhibit a collective excitation nature, wherein excitations are characterized as weighted combinations of Kohn-Sham transitions between occupied and virtual molecular orbitals. 
Transition weights for important excited states near the plasmon energies are listed in the Supplementary Material (SM), Table S1.
From the table it is evident that the states consist of several transitions with similar weights. 
This multi-reference character of the plasmonic excited states has been reported previously.\cite{DFTB-PDC-2022-Zhang, DFTB-PDC-2023-Zhang} \\

The plasmon energy for the individual nanorods shows a noticeable red shift with increasing size of the particle, Fig.\ref{sys_fig}(b), such that
for particles consisting of 73, 97, 121, 133, and 145 atoms, the plasmon energies peak at 1.57, 1.45, 1.34, 1.27, and 1.25 eV, respectively.
Indeed we find a linear decrease in plasmon energies with particle size as shown in the SM Fig.S1(a).
This spectral shift with the nanorod length is consistent with previous experimental\cite{Rod-Plasmon-2000-Sayed, Expt-Rod-2005-Li, Gold-Nanorod-2014-Schatz, Rod-2024-Jin} and theoretical\cite{Theory-Rod-1999-Link, TDDFT-Rod-2014-Weissker} studies. 
Moreover, the absorption amplitude increases with the size of the particle due to the increased transition dipole and density of states accessible for larger particles. \\

The absorption spectra of the dimer systems are influenced by the individual absorption characteristics of each particle and their interactions.
However, the absorption is predominantly influenced by the larger particles in the dimer structure due to the larger absorption cross-sections compared to the smaller ones, Fig.\ref{sys_fig}(c). 
The interaction between the particles causes a slight shift in the plasmon energy of the larger particle towards lower energies.
This shift becomes more pronounced as the distance between the particles decreases, with the peak shifting by up to 25 meV for an inter-nanoparticle spacing of 0.6 nm.
However, we note that the spectral splitting is smaller when there are significant differences between the plasmon energies of the two particles in the dimer systems.
On the other hand, dimers composed of particles of equal length exhibit a larger energy shift, Fig.S1(b).
Similar to single nanorod systems, the plasmon peaks show redshift as the length of the particle increases. 
The plasmon energies for the longitudinal mode of the dimer systems are provided in Table I.
The trends in the spectral shifts were also observed in previous studies.\cite{Rod-Dimer-2009-Schatz, Rod-2013-Mulvaney, Rod-Dimer-2014-Zhu, Rod-Dimer-2022-Alsawafta} 
Furthermore, we observe transverse modes from the $XX$ and $YY$ polarizability components, which appear at relatively higher energies, near 2.05 eV, as depicted in Fig.S5. 
This mode does not show an energy shift with the nanoparticle length as we maintain a fixed width for the particles. \\

\begin{figure}[h]
    \centering
    \includegraphics[width = 0.45 \textwidth]{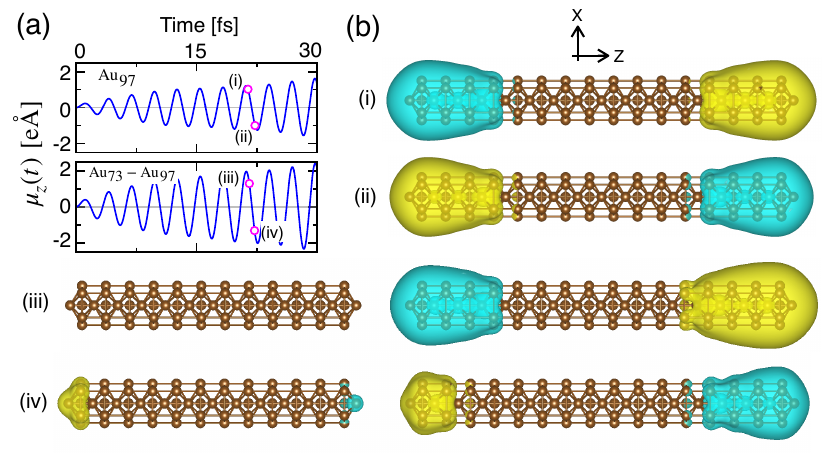}
    \caption{Snapshots of charge density difference (CDD) for Au$_{97}$ and Au$_{73}$-Au$_{97}$ systems interacting with a CW laser. 
    (a) Induced dipole and (b) CDD for selected points highlighted in the induced dipole plots.
    Here, driving frequency $\omega_p=1.45$ eV, peak intensity $I_p=1.3\times10^{7}$ W/cm$^2$, and polarization direction is along the $Z$ axis.
    Yellow color represents positive and cyan color represents negative surfaces with equal absolute amplitudes.}  
   \label{charge_density_fig}
\end{figure}

In Figure \ref{charge_density_fig}, we display the induced dipole and charge oscillation for Au$_{97}$ and Au$_{73}$-Au$_{97}$ systems interacting with a CW laser of form $E_z(t)=\hat{e}_zE_p\sin(\omega_pt)$. 
The central frequency of the laser is tuned to match the respective plasmon energies of the particles. 
The induced dipole for both systems oscillates around their mean values, with its amplitude increasing as more energy is transferred to the systems through interaction with the field over time.
We select two points along the time axis representing opposite directions for dipoles but with nearly equal absolute amplitude. 
For these selected points, charge density difference (CDD) surfaces for both systems are shown in Fig.\ref{charge_density_fig}(b). 
The CDD is calculated as $\Delta q(t)=q(t)-q(0)$, where $q(0)$ is the initial charge density. 
The surface changes sign when the dipole changes its direction, clearly demonstrating the oscillation of induced charge due to plasmon excitation.
Interestingly, the absolute amplitude of the surface is symmetric for the single particle around the center of the nanorod; however, this symmetry is disrupted in dimer geometries in the presence of a second nanoparticle. 
This asymmetric charge oscillation impacts the SHG process, which we discuss in the following. 

\subsection*{B. Second and Third Harmonic Generation (SHG and THG)}
\noindent In this section, we discuss harmonic generation (HG) processes for nanorod systems and their dependency on pulse parameters and nanoparticle geometries. 
Signals are obtained following the Eq.\ref{hg_eq} through the computation of induced dipoles.
Here we report signals obtained by integrating dipoles from all possible directions as given in Eq.\ref{hg_eq}, however, the rod structure results are dominated by dipoles along the rod axis.
Throughout we use the following laser parameters unless otherwise specified: 
longitudinal excitation of the plasmon mode achieved by suitably selecting a driving frequency depending on the system, a pulse duration $T_p$ of 14 fs, a filter function (damping) width $\tau$ of 30 fs, a laser polarization direction of $Z$, and an inter-particle spacing $h$ of 0.6 nm.  

\begin{figure}[h]
    \centering
    \includegraphics[width = 0.4 \textwidth]{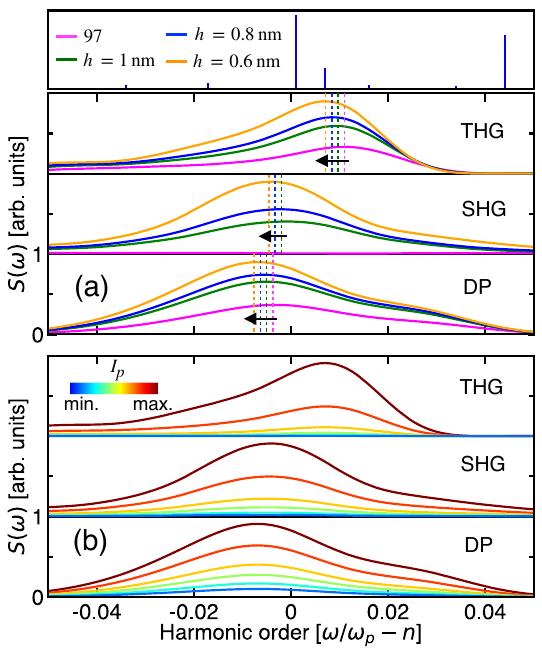}
    \caption{(a) Fundamental dipole (DP), second harmonic generation (SHG), and third harmonic generation (THG) spectra for single Au$_{97}$ particle and Au$_{73}$-Au$_{97}$ dimers with various inter-particle spacing $h$.
    Here, $n$ is the order of the induced optical process, $n=$ 1, 2, and 3 for DP, SHG and THG processes, respectively. 
    In the top row we display oscillator strengths for Au$_{97}$ shifting the energy of each state by $\omega_p$ to the lower energy. 
    Here the pump peak intensity is $I_p=2.5\,\times\,10^8$ W/cm$^2$, and driving frequency $\omega_p=1.45$ eV.
    For comparison, all the spectra are normalized to the corresponding spectra of dimer systems for $h=0.6$ nm. Vertical dashed lines and a black arrow, following the same color scheme as the original spectra, indicate the red shift of spectra as the inter-particle spacing decreases.
    (b) DP, SHG, and THG spectra for different field intensities. 
    Here, the field intensity increases from $2.3\,\times\,10^7$ to $2.5\times10^8$ W/cm$^2$ as indicated by the color gradient from blue to red and $\omega_p=1.45$ eV.
    The vertical dashed gray line represents the central laser frequency.  
    For comparison, all the spectra are normalized to the corresponding spectra for maximum peak intensity.}  
   \label{hg_spec_fig}
\end{figure}

\subsubsection*{1. End-To-End Nanorod Dimer} 
\noindent Figure \ref{hg_spec_fig}(a) illustrates fundamental dipole (DP), second harmonic generation (SHG), and third harmonic generation (THG) spectra for Au$_{73}$-Au$_{97}$ systems with different gaps $h$ between the particles and a single Au$_{97}$ nanorod. 
Nonlinear polarization is induced when particles are excited at 1.45 eV, very close to their plasmon energy, through interaction with an intense laser pulse of the form given in Eq.\ref{pulse_eq}.
For both single and dimer systems, the DP and THG spectra, obtained through the fast Fourier transformation of the induced dipole, peak around their respective energies i.e., around $\omega_p$ for DP and $3\times\omega_p$ for THG. 
However, only the dimer systems lacking the longitudinal symmetry exhibit activity in SHG around the energy $2\times\omega_p$, while the single particle provides negligible amplitude for SHG, as clearly evident in Fig.\ref{hg_spec_fig}(a).
Further, we refer to the SM, Fig.S2(a), for the comparison of full HG spectra for dimer systems with signals from each nanorod systems. 
As previously noted, unlike dimer systems of unequal lengths, a single nanorod possesses inversion symmetry along the longitudinal direction and hence remains inactive in SHG. \\

The first harmonic (DP) spectra contain two major peaks around the central laser frequency as observed in Fig.\ref{hg_spec_fig}(a).  
These two peaks correspond to excited states with significant oscillator strengths that appear within the energy window of the pump laser as observed in the top row of the figure. 
The peak position of SHG spectra (only for dimer systems) follows the peaks of DP spectra, while the higher energy peak amplitude is significantly suppressed compared to the DP spectra. 
However, THG spectra show maxima at higher energy that is in between the two peaks in the DP results, which is suggestive of combination mode excitation. 
Overall, the spectrum gets narrower as we move to higher order processes from DP to THG as the higher order processes are more sensitive to the resonant driving field. \\

Moreover, the DP spectra in Fig.\ref{hg_spec_fig}(a) show a slight red-shift with increased peak amplitudes as $h$ decreases, a trend consistent with the red-shift observed in the linear absorption spectrum, as shown in Fig.\ref{sys_fig}(c). 
Interestingly, both SHG and THG spectra also follow a similar energy shift towards lower energies with decreasing $h$.
The spectral shifts in the SHG and THG spectra are comparable to the shift of the dipole spectra.  
Additionally, the amplitudes of the nonlinear signals increase similarly to the DP response but at a higher rate, as depicted in Fig.S2(b). 
Similar enhancements in SHG\cite{SHG-Dimer-2012-Roch} and THG\cite{THG-Dimer-2007-Novotny, Nonlinear-Gold-Nanowire-2012-Novotny} results for smaller inter-particle spacings have also been reported in experiments.

\begin{figure}[h]
    \centering
    \includegraphics[width = 0.4 \textwidth]{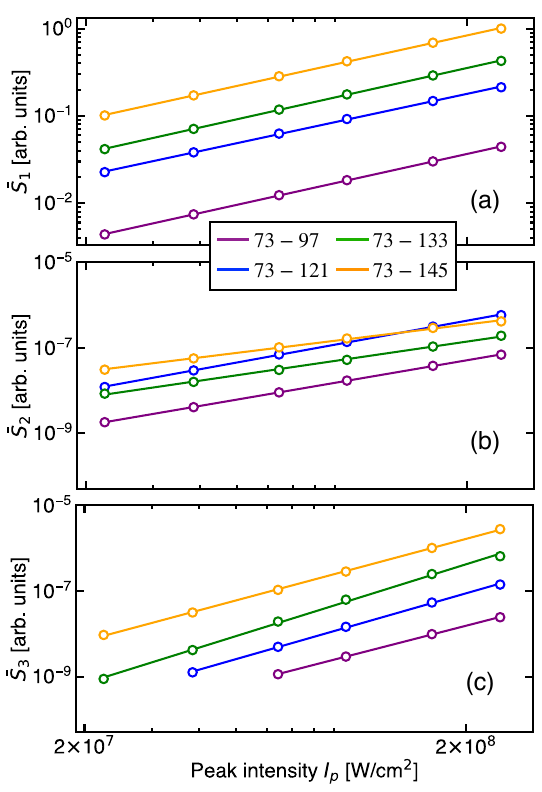}
    \caption{Integrated (a) DP, (b) SHG, and (c) THG signals as a function of pump intensity for all dimer systems used in this study. 
    All the signal amplitudes are normalized by a constant factor for comparison.}  
   \label{hg_int_fig}
\end{figure}

\begin{table}[h]
\centering
\label{hg_int_table}
\begin{threeparttable}
        \caption{Length ratio, LSPR energy, and exponents for power scaling of the DP, SHG, and THG processes as a function pump intensity for dimer systems. 
    }
    \renewcommand{\arraystretch}{1.4}
    \begin{tabular}{| c | c | c | c | c | c |} 
        \hline
         System & Length Ratio ($l_2/l_1$)& LSPR [eV] & DP  & SHG & THG \\
         \hline
         \hline
             73-97 & 1.33 & 1.45 & 0.97 & 1.52 & 2.29 \\
             73-121 & 1.66 & 1.34 & 0.94 & 1.62 & 2.54 \\
             73-133 & 1.83 & 1.27 & 0.97 & 1.3 & 2.78 \\
             73-145 & 2 & 1.24 & 0.96 & 1.1 & 2.38 \\
         \hline
    \end{tabular}
\end{threeparttable}
\end{table}

\subsubsection*{2. Pump Intensity}
\noindent The DP and HG spectra for different pump intensities for the Au$_{73}$-Au$_{97}$ system are shown in Fig.\ref{hg_spec_fig}(b).
The overall amplitude for all spectra increases with the field intensity as a higher photon density induces a larger polarization within the nanorod systems.
Importantly, despite the increase in amplitude, the shape of the spectra remains largely unchanged across all the intensities. 
This highlights the stability of the process under varying pump intensities. \\

The scaling of the DP and HG signals as a function of pump intensity for all the dimer systems is shown in Fig.\ref{hg_int_fig}.
Here, we use integrated spectra for the yield of the signals as $\bar{S}_{n}=\int_{\Delta\omega_{n-}}^{\Delta\omega_{n+}}\d\omega S(\omega)$, where $\Delta\omega_n$ is the energy window around the center of the $n$-th order signal.  
Throughout this paper, we maintain a consistent energy window size, defined as $\Delta\omega_{n\pm}=(n\pm0.05)\times\omega_p$. \\

The comparison of signal yield for different pump intensities reveals significant differences among the various HG processes.
As observed in the Fig.\ref{hg_int_fig}, all the signals follow the power law dependency, expressed as $\bar{S}_n\propto I_p^n$, with pump intensity $I_p$.
The slopes obtained from linear fits represent the power dependence of the respective processes on the incident field intensity.
All the slopes, along with their respective plasmon energies, are listed in Table I.
As is evident from the table, the DP signals depend of the pump intensity almost linearly with an exponent between $0.94-1$ as expected.
On the other hand, the slopes for the SHG signals range between 1.1 and 1.6, while those for the THG signals are between 2.3 and 2.8.
This observation reflects the nonlinearity in SHG and THG compared to the DP signals. 
Moreover, the deviation of the HG exponents from the perturbative results ($n=2$ for SHG and $n=3$ for THG) indicate a saturation phenomenon, signifying the non-perturbative characteristics of HG under intense laser pulses.
Deviation from the perturbative results for the lower-order harmonics (SHG and THG) has been noted in previous works.\cite{THG-2020-Wang, HG-2022-Kivshar, HG-2023-Meier} 
Specifically, exponents for the power law dependencies on-field intensity were found to be smaller than those predicted by perturbation theory. 
Furthermore, we note contributions from the overlap between the dominant dipole and HG signals, leading to significantly smaller SHG and THG exponents. 
This effect is particularly pronounced for larger dimer systems, where the amplitudes of the DP signals are substantially larger. \\

Detectable HG signals are observed only when the nanorod systems interact with intense laser fields, while dipole modes at the fundamental frequency can be excited at relatively weak field intensities. 
As a result, the amplitudes of the dipole signals are significantly larger than those of the HG by several orders of magnitude, as observed in Fig.\ref{hg_int_fig}(a). 
Indeed, the amplitudes of SHG and THG signals are comparable within the same orders of magnitude. 
Particularly, SHG signals exhibit larger amplitudes at lower intensities compared to THG signals, while THG signals become dominant at higher intensities, given their higher sensitivity to field intensities. \\

The amplitudes of DP and THG signals increase with increasing nanorod lengths from 73-97 to 73-145, Fig.\ref{hg_int_fig}(a, c). 
A similar trend is also observed for SHG except for the 73-121 system, which has a higher SHG yield for an intermediate length. 
One possible reason for this dependence could be the presence of a larger number of states with significant oscillator strength that are near the plasmon energy (Fig. \ref{sys_fig}(b, c)) than occurs for the other nanorods, which serves to amplify the SHG signal.
This effect does not extend to the DP and THG results, so asymmetry in the excited states is also a factor.
We note experimental studies that have reported an increase in the SHG\cite{SHG-Rod-2023-Chandra, Rod-SHG-2016-Ledoux-Rak} and THG\cite{THG-Plasmon-2005-Orrit} yields with the increasing nanoparticle size. 
Our results show a similar trend, particularly for larger particles, where the oscillator strengths for plasmon-excited states dominate significantly over other contributing states that may overlap with plasmon excitation.  For nanorods, one experiment shows an increase in SHG with an increasing aspect ratio\cite{SHG-Rod-2023-Chandra} while another shows a decrease\cite{Rod-SHG-2016-Ledoux-Rak}. This is related to the variations with length that we see in Fig. \ref{sys_fig}(b)) 

\begin{figure}[h]
    \centering
    \includegraphics[width = 0.4 \textwidth]{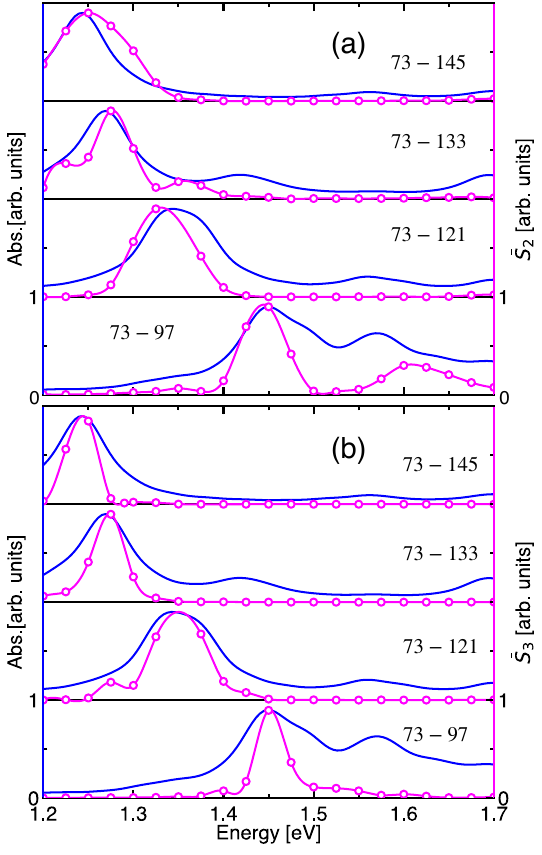}
    \caption{Pump frequency dependent (a) SHG and (b) THG signals.
    HG signals are compared with the corresponding linear absorption spectra.
    For comparison, peak amplitudes for all the spectra are normalized.
    Here, $I_p=9.7\times10^7$ W/cm$^2$.}  
   \label{hg_wp_fig}
\end{figure}

\subsubsection*{3. Pump Frequency}
\noindent Figure \ref{hg_wp_fig} displays SHG and THG signals as a function of pump frequency (i.e., action spectra) for all the dimer systems used in this study.
As discussed earlier, the LSPR energy undergoes a significant red shift with increasing rod length. As a result,  
analyzing how incident frequency affects HG processes by changing the particle length and hence modifying the LSPR energy reveals the important role of plasmon resonance in HG efficiency. \\

Irrespective of the nanorod lengths, both SHG and THG signals are similar to the linear absorption spectrum of the nanorod systems, Fig.\ref{hg_wp_fig}.
However, the HG signals are significantly suppressed outside a region with a width $\sim\,0.1$ eV that corresponds to LSPR excitation.
Although smaller transitions exist in the linear spectra outside this LSPR energy range, these smaller absorption amplitudes are not effective in inducing the nonlinear processes. 
This clearly demonstrates the role of the LSPR in HG processes. 
Previous studies  (both theory and experiment) have reported a similar frequency dependence for both SHG and THG processes where the nonlinear responses closely follow plasmon excitation.\cite{Hyper-Rayleigh-2002-Schtaz, SHG-Plasmon-2007-Fort, THG-WP-2018-Hentschel, SHG-Wp-Polarization-2020-Sukharev} \\ 

Furthermore, the THG signals exhibit narrower profiles compared to SHG signals.
Interestingly, we observe that the THG maxima align closer with the linear absorption spectra peaks compared to SHG. 
This difference between THG and SHG can be attributed to the higher sensitivity of THG to the incident field amplitude relative to SHG.

\begin{figure}[h]
    \centering
    \includegraphics[width = 0.4 \textwidth]{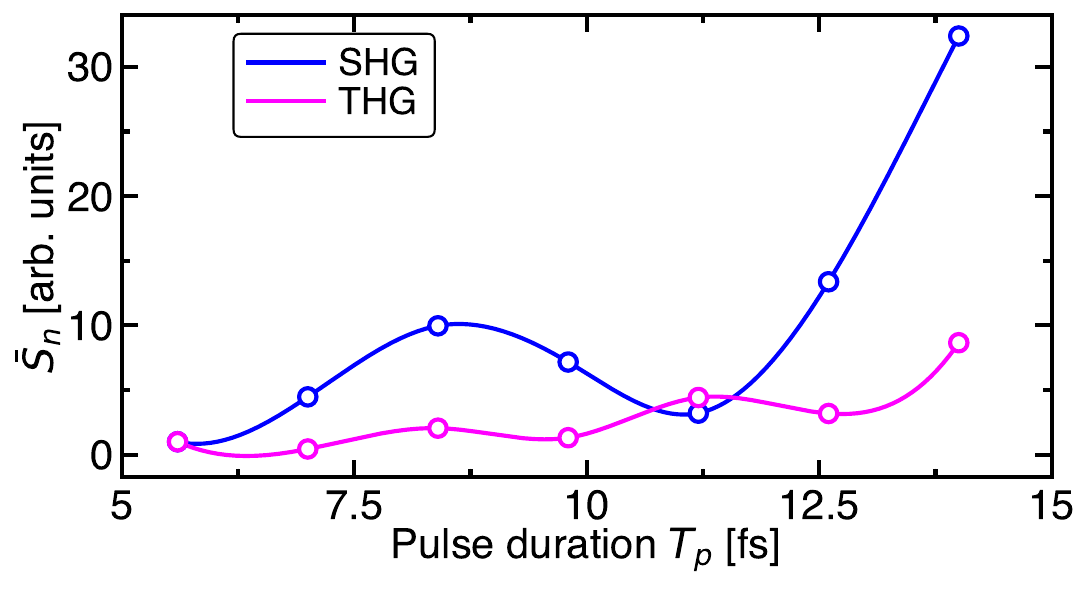}
    \caption{Integrated SHG and THG signals as a function of pulse duration $T_p$ for Au$_{73}$-Au$_{97}$ systems.
    For comparison, all signals are normalized to 1 for the smallest $T_p$ value.
    Here, $I_p=9.7\times10^7$ W/cm$^2$.}  
   \label{hg_tp_fig}
\end{figure}

\begin{figure}[h]
    \centering
    \includegraphics[width = 0.4 \textwidth]{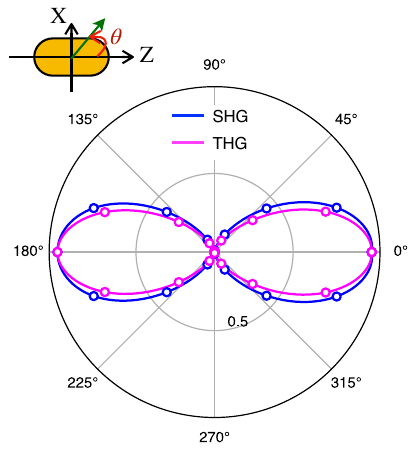}
    \caption{Integrated SHG and THG signals as a function of pump laser polarization direction for Au$_{73}$-Au$_{97}$ system, making an angle $\theta$ with respect to the $Z$ axis in the $X-Z$ plane.
    Here, $\theta=0^0$ signifies alignment of the laser polarization with the $Z$ axis as shown in the inset and $I_p=9.7\times10^7$ W/cm$^2$.
    The peak amplitudes ($\theta=0^0$) for both SHG and THG are normalized to 1.}  
   \label{shg_polar_fig}
\end{figure}

\subsubsection*{4. Pump Duration}
\noindent The dependence of HG signals on laser pulse duration $T_p$ is shown in the Figure \ref{hg_tp_fig}.
This study explores how variations in the temporal characteristics of the laser pulse influence the induced signals, providing insights into the intricate dynamics of the nonlinear optical processes under consideration. \\

Pulse duration is changed in such as a way that the energy content of the pulse remains unchanged, i.e. $\int_{-\infty}^{\infty}\d tE^2(t)=$Constant.
Therefore, the peak intensity of the pulse decreases as $T_p$ increases. 
We note that the pulse always starts at $t=0$ and hence the center of the pulse shifts to larger values with the increase of $T_p$. \\

Interestingly, the overall amplitude for both SHG and THG signals tends to increase on average (with oscillations) with the increase in $T_p$, and the SHG amplitude rises rapidly above 12 fs.
The increase suggests that longer pulses, for which there is dephasing of the electronic states while they are being excited are more efficient for harmonic generation.  
This can result because the dephased states have higher oscillator strength. 
The results in Figure \ref{hg_tp_fig} are consistent with previous experiments where several detectable harmonics are observed for long pulse durations but were not observed for short pulse durations.\cite{HG-2022-Kivshar}

\subsubsection*{5. Pump Polarization}
\noindent Laser polarization dependent HG signals for Au$_{73}$-Au$_{97}$ systems are shown in Figure \ref{shg_polar_fig}.
As seen in the figure, maximum yield is obtained when the laser polarization aligns with the longitudinal axis of the nanorod systems. 
Both SHG and THG amplitudes decrease as the polarization angle changes from the longitudinal alignment.
This observation again confirms that resonant excitation of plasmon modes leads to efficient HG processes.
A similar polarization dependence for SHG has also been reported previously for nanorod systems \cite{HG-Wp-Polarization-2021-Sukharev} as well as for nanosphere dimer systems \cite{SHG-Dimer-2012-Roch}.
However, THG amplitude decreases more rapidly compared to SHG due to its higher sensitivity to the laser pulse.\\

We do not observe significant amplitudes for perpendicular alignment of laser polarization with the longitudinal axis, even when changing the pump energy up to 2.05 eV, where the transverse mode peak appears. 
However, the $XX$ and $YY$ polarizability components are significantly smaller than the $ZZ$ component near the transverse energy, Table II and Fig.S5. 
This is because the nanorod is significantly narrower and contains only a few atoms along the transverse direction for the systems we studied here.\\

\begin{table}[h]
\centering
\begin{threeparttable}
    \label{pol_table}
    \caption{Imaginary part of $XX$, $YY$, and $ZZ$ components for $\alpha$, $ZZZ$ component for $\beta$, and $ZZZZ$ for $\gamma$.
    The results are obtained near the respective longitudinal plasmon peak energies for $ZZ$ component of $\alpha$, $ZZZ$ component of $\beta$, and $ZZZZ$ component of $\gamma$. 
    For the remaining $\alpha$ components, they are evaluated around the transverse plasmon energy.
    $\alpha$ values are obtained through Eq.\ref{alpha_eq}, while $\beta$ and $\gamma$ are obtained through dipole fitting following the work by Li and co-workers \cite{RT-HyperPol-2013-Li}.
    }
    \renewcommand{\arraystretch}{1.4}
    \begin{tabular}{|c|c|c|c|c|c|c|}
        \hline
         \multirow{2}{*}{Systems} & \multicolumn{3}{c|}{$\alpha$ [cm$^3$]} & $\beta_{zzz}$ & $\gamma_{zzzz}$\\
         \cline{2-4}
    & $\alpha_{xx}$ & $\alpha_{yy}$ & $\alpha_{zz}$ & [10$^{-26}$ esu]& [$10^{10}$ a.u.] \\
    & \big[10$^{-22}$\big] & \big[10$^{-22}$\big] & \big[10$^{-20}$\big] &  &  \\
         \hline
         \hline
         73-97 &1.29&1.29&2.34& 2.1 & - \\
         73-121 &1.47&1.47&4.89& 2.2 & 8 \\ 
         73-133 &1.51&1.51&5.94& 3.2 & 141.3 \\ 
         73-145 &1.75&1.75&9.76& 3 & 256.5 \\ 
         \hline
    \end{tabular}
  \end{threeparttable}
\end{table}

\subsubsection*{6. (Hyper-)Polarizability Calculations}
\noindent To quantify our study of nonlinear optical properties, we have extracted the polarizability ($\alpha$), first hyper-polarizability ($\beta$), and second hyper-polarizability ($\gamma$).
Table II presents the $XX$, $YY$, and $ZZ$ components of $\alpha$, $ZZZ$ component of $\beta$, and $ZZZZ$ component of $\gamma$.
While polarizabilities are obtained through Eq.\ref{alpha_eq}, hyper-polarizabilities are obtained from dipole fitting following the method introduced by Li and co-workers\cite{RT-HyperPol-2013-Li}.
Furthermore, we computed polarizabilities in the frequency domain by evaluating excited states and transition dipoles between ground and excited states. 
These results exhibit agreement with each other for both peak positions and amplitudes, as shown in Fig.S4.
Note that we only provide $ZZZ$ and $ZZZZ$ components for $\beta$ and $\gamma$, respectively, as the transverse excitation is negligibly small for the HG signals for the systems we consider. 
For the $ZZ$, $ZZZ$, and $ZZZZ$ components, results are provided at the respective plasmon energies of the longitudinal mode. 
On the other hand, the results for the $XX$ and $YY$ components of $\alpha$ are provided at the transverse plasmon energy of 2.05 eV. \\

We note that because of the high density of electronic states that are close to resonance with the driving field, a different pulse shape for the driving field is employed in this study compared to the form used in the study by Li and co-workers\cite{RT-HyperPol-2013-Li}.  
The use of Eq.\ref{pulse_eq} involves a more gradual increase in the driving field than the linear ramp over one optical period that was used by Li and co-workers.
For comparison of our method with reported results by Li and co-workers, we studied the same molecule, para-nitroaniline (pNA), and the resulting (hyper-)polarizabilities are given in Tables S2.
The fitting performance for that molecule and for one of the nanoparticles we studied is illustrated in Fig.S7.
Overall the agreement with Li and colleagues is quite good. 
For further comparison, we computed frequency-dependent (hyper-)polarizabilities for the pNA molecule and compared them with previous theoretical and experimental results; see Figures S6 and S8, as well as Tables S3 for details.\\

As observed in Table II, both polarizability and hyper-polarizabilities increase with size of the system. 
Furthermore, our findings are consistent with reported linear absorption results that utilize the TD-DFTB method for icosahedral Ag and Au nanoparticles,\cite{RT-TDDFTB-Hot-2019-Sanchez} with absorption cross sections in the range of $10^{-15}$ to $10^{-14}$ cm$^{2}$ for both their results and ours.
However, it is also important to note that the polarizability results are sensitive to the choice of damping rate. 
While Douglas-Gallardo et al.\cite{RT-TDDFTB-Hot-2019-Sanchez} employed a damping time of 7 fs, we find that for this choice of damping time, polarizabilities are nearly a factor of 2 smaller than what we obtain for 20 fs, as shown in Fig.S3. \\

Concerning the first hyper-polarizabilities, Ngo et al.\cite{Rod-SHG-2016-Ledoux-Rak} reported measured $\beta$ values for nanosphere and nanorod systems with sphere diameters down to 3 nm, and the results for the smallest particles are similar in magnitude $10^{-26}-10^{-25}$ esu to our results.  
We also note that there have been previous calculations of static hyper-polarizabilities using TDDFT for gold clusters, and the results are much smaller, $10^{-29}$ esu, as expected.\cite{HG-2022-Liang}  \\ 

For the second hyper-polarizabilities, although there have been several reported THG measurements for gold nanoparticles (usually as films) as has been reviewed,\cite{Plasmon-2018-Lippitz} we are not aware that values of $\gamma$ have been reported. 
The values we report, roughly $10^{12}$ atomic units, or $5\times10^{-28}$ esu, are larger about a factor of $10^3$ than the largest values tabulated for optimized organic oligomers.\cite{THG-2998-Gunter} 
This seems reasonable given that we are calculating the resonant response rather the the nonresonant response that was considered for the organic molecules. 

\section{Conclusions}
\noindent In conclusion, this study investigates the dependence of laser pulse parameters on nonlinear optical processes, particularly second and third harmonic generation, for end-to-end Au nanorod dimer systems for variable length ratios and inter-particle spacing. 
All harmonic generation (HG) signals are derived from induced dipoles for nanorod systems interacting with external laser pulses computed using the real-time time-dependent density functional tight-binding (RT-TDDFTB) method. 
The characteristics of the obtained HG signals align with previously reported results based on TD-DFT, but the calculations are orders of magnitude faster, which means we can study much larger structures. \\

Detectable second harmonic generation (SHG) signals are observed only in nanorod dimer systems with broken longitudinal symmetry due to different length ratios. 
The HG yield increases with the particle size, and a smaller inter-particle spacing is found to be more effective for these processes to occur.
The intensity dependence of all generated signals follows a power law in the electric field intensity, with exponents comparable to perturbative results, although higher order harmonics involve non-perturbative contributions. 
The study of HG yield as a function of incident frequency and laser polarization highlights the significance of resonant excitation of plasmon modes for efficient HG processes. 
Longer pulse durations are identified as producing more efficient nonlinear excitation for a fixed energy per pulse.
Finally, we report the polarizability and first and second hyper-polarizability components for the dimer systems.
These are computed from the induced real-time dipoles using a Dirac delta excitation for the polarizabililties and using pulsed excitation for the hyper-polarizabilities.  The calculated first hyper-polarizability compares well with a measured result for a small gold nanorod. \\

Overall, these findings highlight the use of the RT-TDDFTB method in elucidating the role of laser parameters and nanorod characteristics in governing nonlinear optical processes. 
This study provides valuable insights for the design and optimization of nanoscale optical devices.

\section*{SUPPLEMENTARY MATERIAL}

The supplementary material includes details of the frequency domain polarizability, plasmon excited states, harmonic generation spectra, and (hyper-)polarizability results for the para-nitroaniline molecule.

\section*{ACKNOWLEDGMENTS}
\noindent This research was supported by the Office of Basic Energy Science, Department of Energy, through grant DE-SC0004752.

\section*{AUTHOR DECLARATIONS}

\subsection*{Conflict of Interest}
The authors have no conflicts to disclose.

\section*{References}

\bibliography{draft.bib}

\end{document}


\begin{center}\bfseries\Large
Supplementary Material: \\
Laser Pulse Induced Second- and Third-Harmonic Generation of Gold Nanorods with Real-Time Time-Dependent Density Functional Tight Binding (RT-TDDFTB) Method
\\ [7mm]
\rm\large
Sajal Kumar Giri\footnote{sajal.giri@northwestern.edu} and 
George C. Schatz\footnote{g-schatz@northwestern.edu} \\ [2mm]
\small{Department of Chemistry, Northwestern University, 2145 Sheridan Road, Evanston, Illinois 60208, United States}
\end{center}

\vspace{0.2cm}


\tableofcontents
\thispagestyle{plain}

\clearpage

\section{Supplementary Note}
The frquency domain polarizability  is computed as\cite{Book-2002-Schatz} 
\begin{equation}
    \alpha_{ab}(\omega)\=\sum_{n>0}\Bigg{[}\frac{\langle\,0|\hat{\mu}_a|n\rangle\langle\,n|\hat{\mu}_b|0\rangle}{\omega_{n0}-\omega-\i\gamma/2}+
    \frac{\langle\,0|\hat{\mu}_b|n\rangle\langle\,n|\hat{\mu}_a|0\rangle}{\omega_{n0}+\omega+\i\gamma/2}\Bigg{]}
    \label{pol_freq_eq}
\end{equation}
where $\hat{\mu}_a$ denotes the dipole momnent operator along the axis $a$ and $\omega_{n0}\=\omega_n-\omega_0$ with $n$-th excited and ground state energies $\omega_n$ and $\omega_0$, respectively. 
\section{Plasmon Excited States}
\begin{table}[h]
    \centering
    \begin{threeparttable}
            \caption{Transition weights for Au nanorod systems from the LR-TDDFTB calculation.
        }
        \renewcommand{\arraystretch}{1.2}
        \begin{tabular}{| c | c | c | c | c | c | c | c |} 
            \hline
             \multicolumn{2}{|c|}{Au$_{73}$} &  \multicolumn{2}{c|}{Au$_{97}$} & \multicolumn{2}{c|}{Au$_{121}$} & \multicolumn{2}{c|}{Au$_{145}$}\\
             \hline
             Transition & Weight & Transition & Weight & Transition & Weight  & Transition & Weight \\
             \hline\hline
                  357   $\rightarrow$   407  &   0.5217 &  495   $\rightarrow$   540 & 0.7734 &   671   $\rightarrow$   735   & 0.933 &  751   $\rightarrow$   807   &  0.8038 \\
                  356   $\rightarrow$   408  &   0.5217 &  485   $\rightarrow$   537 & 0.431 &   674   $\rightarrow$   734   &  0.1881 &   795   $\rightarrow$   806  &   0.2428 \\
                  364   $\rightarrow$   406  &   0.2751 &  486   $\rightarrow$   538 & 0.4304 &   675   $\rightarrow$   733   &  0.1881 &   794   $\rightarrow$   805   &  0.2428 \\
                  367   $\rightarrow$   408  &   0.2535 &  488   $\rightarrow$   541 & 0.0786 &   674   $\rightarrow$   733   &  0.1131 &   733   $\rightarrow$   798   &  0.2334 \\
                  368   $\rightarrow$   407  &   0.2535 &  487   $\rightarrow$   542 & 0.0786 &   675   $\rightarrow$   734   &  0.1131 &  734   $\rightarrow$   799    & 0.2333 \\
                  337   $\rightarrow$   407  &   0.2135 &  496   $\rightarrow$   542 & 0.0512 &   730   $\rightarrow$   742  &  0.0727  & 791   $\rightarrow$   801    & 0.1829 \\
                  336   $\rightarrow$   408  &   0.2135 &  497   $\rightarrow$   541 & 0.0512 &   676   $\rightarrow$   734  &   0.0567 &   795   $\rightarrow$   827   &  0.1714 \\
                  365   $\rightarrow$   409  &   0.1756 &  533   $\rightarrow$   540 & 0.045 &   677   $\rightarrow$   733  &   0.0567 &   794   $\rightarrow$   826    & 0.1714 \\
                  366   $\rightarrow$   410  &   0.1756 &  530   $\rightarrow$   542 & 0.0369 &   652   $\rightarrow$   733  &  0.0533 &   791   $\rightarrow$   800    & 0.0869 \\
                  402   $\rightarrow$   411  &   0.1448 &  529   $\rightarrow$   541 & 0.0369 &   651   $\rightarrow$   734   & 0.0533 &   794   $\rightarrow$   827    & 0.0476 \\
                  398   $\rightarrow$   408  &   0.1128 &  533   $\rightarrow$   545 & 0.0328 &   725   $\rightarrow$   734   & 0.0489 &   795   $\rightarrow$   826    & 0.0476 \\
                  399   $\rightarrow$   407  &   0.1128 &  524   $\rightarrow$   549 & 0.0294 &   724   $\rightarrow$   733   & 0.0489 &   791   $\rightarrow$   807    & 0.0465 \\
                  393   $\rightarrow$   412  &   0.0979 &  523   $\rightarrow$   550 & 0.0294 &   730   $\rightarrow$   747   & 0.0392 &   718   $\rightarrow$   805    & 0.0446 \\
                  402   $\rightarrow$   415  &   0.0881 &  522   $\rightarrow$   546 & 0.0279 &   648   $\rightarrow$   733   &  0.0301 &   717   $\rightarrow$   806    & 0.0446 \\
                  399   $\rightarrow$   408  &   0.0703 &  480   $\rightarrow$   541 & 0.024 &   647   $\rightarrow$   734   &  0.0301 &   794   $\rightarrow$   806    & 0.0388 \\
                  398   $\rightarrow$   407  &   0.0703 &  479   $\rightarrow$   542 & 0.024 &   676   $\rightarrow$   733    & 0.0221 &   795   $\rightarrow$   805    & 0.0388 \\
             \hline
    \end{tabular}
  \end{threeparttable}
\end{table}

\begin{figure}[h]
    \centering
    \includegraphics[width = 0.55 \textwidth]{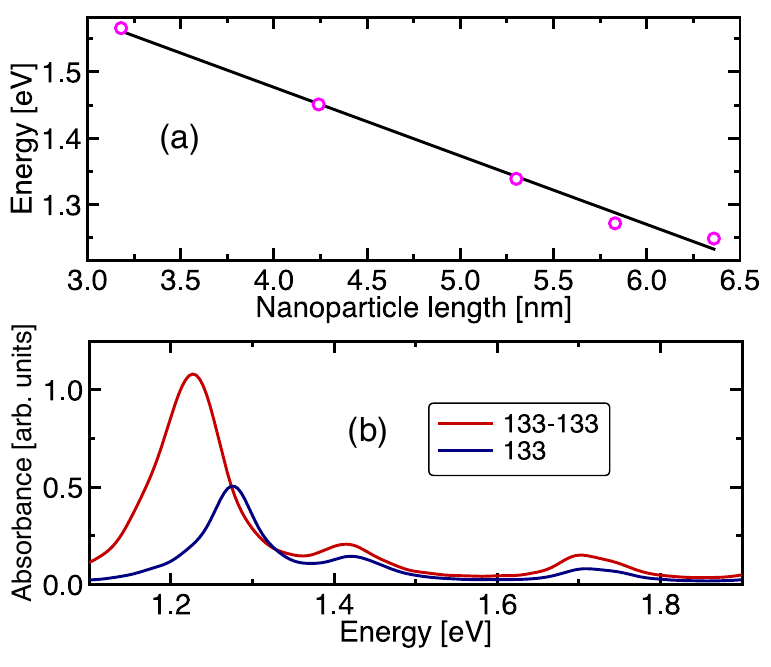}
    \caption{(a) Plasmon energy (longitudinal mode) as a function of nanoparticle length (single particle).
    (b)Absorption spectra for Au$_{133}$ and Au$_{133}$-Au$_{133}$ systems.
    Here, $h=0.6$ nm for the dimer system.}  
   \label{spec_fig}
\end{figure}

\clearpage
\section{Harmonic Generation}
\begin{figure}[h]
    \centering
    \includegraphics[width = 0.55 \textwidth]{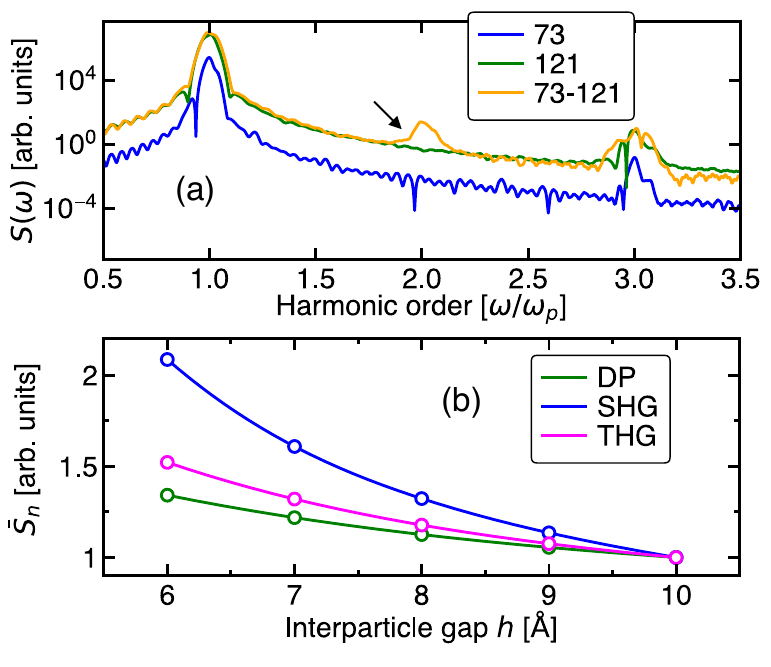}
    \caption{(a)Harmonic signals for Au$_{73}$, Au$_{121}$, and Au$_{73}$-Au$_{121}$ systems.
    Here, $I_p=2.5\times10^8$ W/cm$^2$ and $h=0.6$ nm for the dimer structure.
    (b) Dipole and harmonic signals as a function of inter-particle spacing.
    For comparison, all signals are normalized to 1 for the largest separation considered here, $h=1$ nm.
    Here, $I_p=2.5\times10^8$ W/cm$^2$.}  
   \label{hg_fig}
\end{figure}

\section{(Hyper-)Polarizability Results}

\begin{figure}[h]
    \centering
    \includegraphics[width = 0.55 \textwidth]{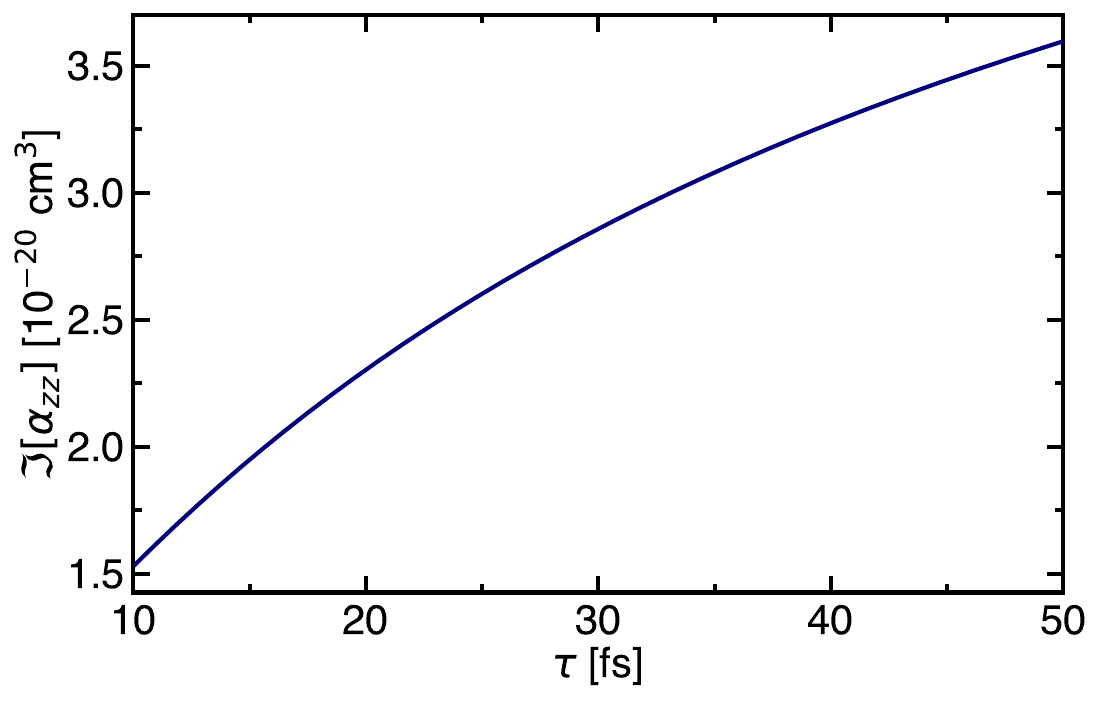}
    \caption{Imaginary part of the $ZZ$ polarizability component at the plasmon energy (longitudinal mode) with the decay time $\tau$ for the Au$_{73}$-Au$_{97}$ system.}  
   \label{alpha_tau_fig}
\end{figure}

\begin{figure}[h]
    \centering
    \includegraphics[width = 0.55 \textwidth]{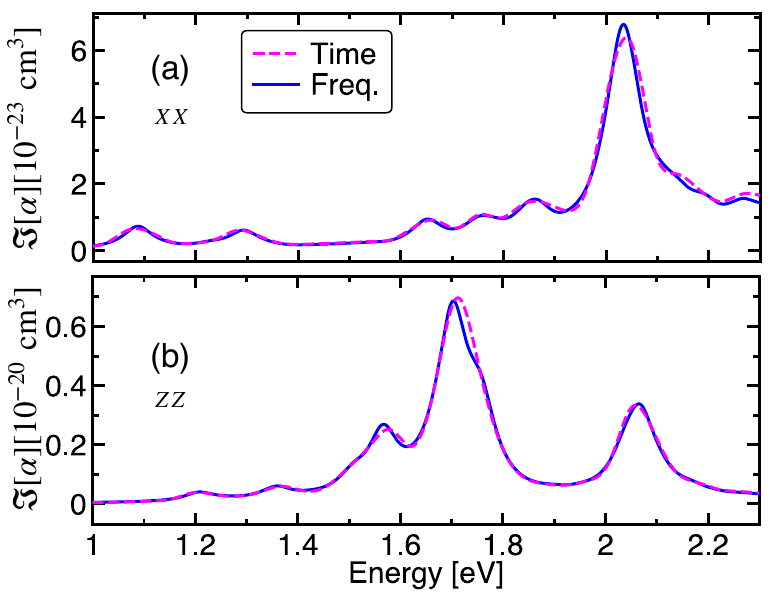}
    \caption{Comparison between time and frequency domain representations for (a) $XX$ and (b) $ZZ$ polarizability components for the Au$_{73}$ system.
    The frquency domain polarizabilites are computed through Eq.\ref{pol_freq_eq} with $\gamma=0.07$ eV.}  
   \label{alpha_time_freq_fig}
\end{figure}

\begin{figure}[h]
    \centering
    \includegraphics[width = 0.55 \textwidth]{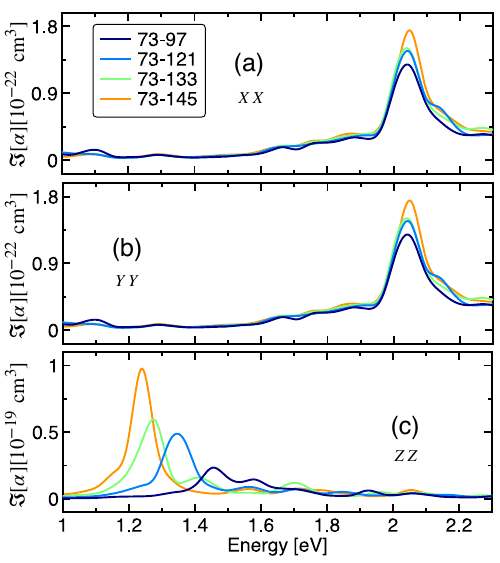}
    \caption{(a) $XX$, (b) $YY$, and (c) $ZZ$ polarizability components for dimer systems with $h=0.6$ nm.}  
   \label{alpha_fig}
\end{figure}

\begin{figure}[h]
    \centering
    \includegraphics[width = 0.55\textwidth]{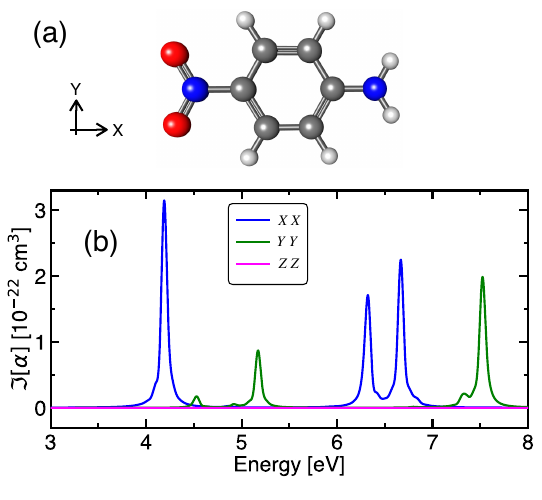}
    \caption{(a) Optimized geometry of the para-nitroaniline molecule. 
    Red: O atom, Blue: N atom, Gray: C atom, and Light gray: H atom.
    (b) Imaginary components of the polarizability for the para-nitroaniline molecule.}
   \label{alpha_pna_fig}
\end{figure}

\begin{figure}[h]
    \centering
    \includegraphics[width = 0.4 \textwidth]{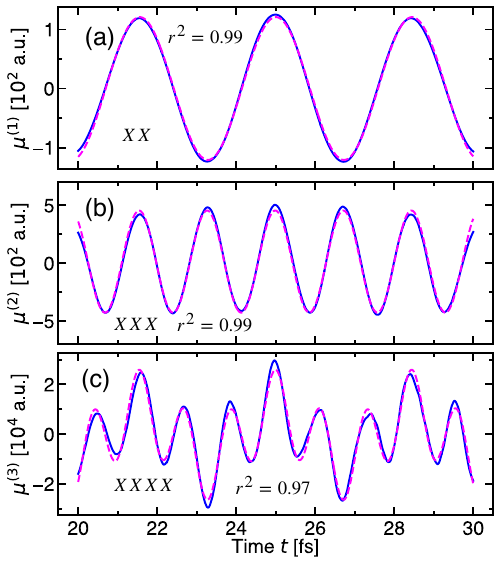}\includegraphics[width = 0.4\textwidth]{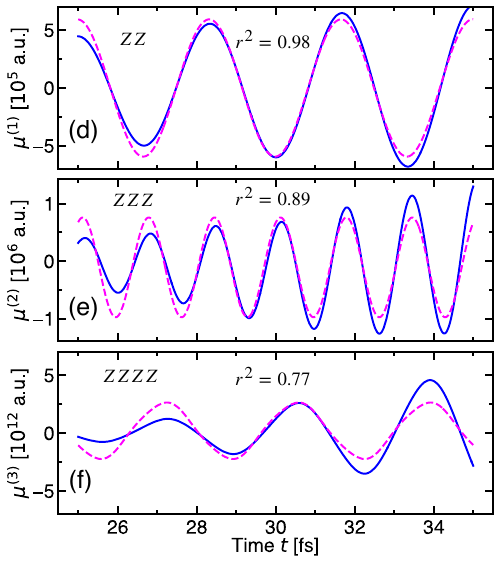}
    \caption{Fitting performed for the dipole components(a, d) $\mu^{(1)}(t)$,  (b, e) $\mu^{(2)}(t)$, and (c, f) $\mu^{(3)}(t)$.
    The left column corresponds to the para-nitroaniline molecule (the same molecules as studied by Li and co-workers \cite{RT-HyperPol-2013-Li}), while the right column represents the Au${73}$-Au${145}$ system.
    The magenta dashed lines represent the fitting, while the blue lines denote the reference dipole components.
    $r^2$ values for the fitting are given in each subplot.
    We used a base intensity of $1.3\times10^{11}$ W/cm$^2$ for the molecule, while for nanoparticles, we used $3\times10^{7}$ W/cm$^2$ for the computation of dipole components.}
   \label{mu_fitting_fig}
\end{figure}

\begin{figure}[h]
    \centering
    \includegraphics[width = 0.55\textwidth]{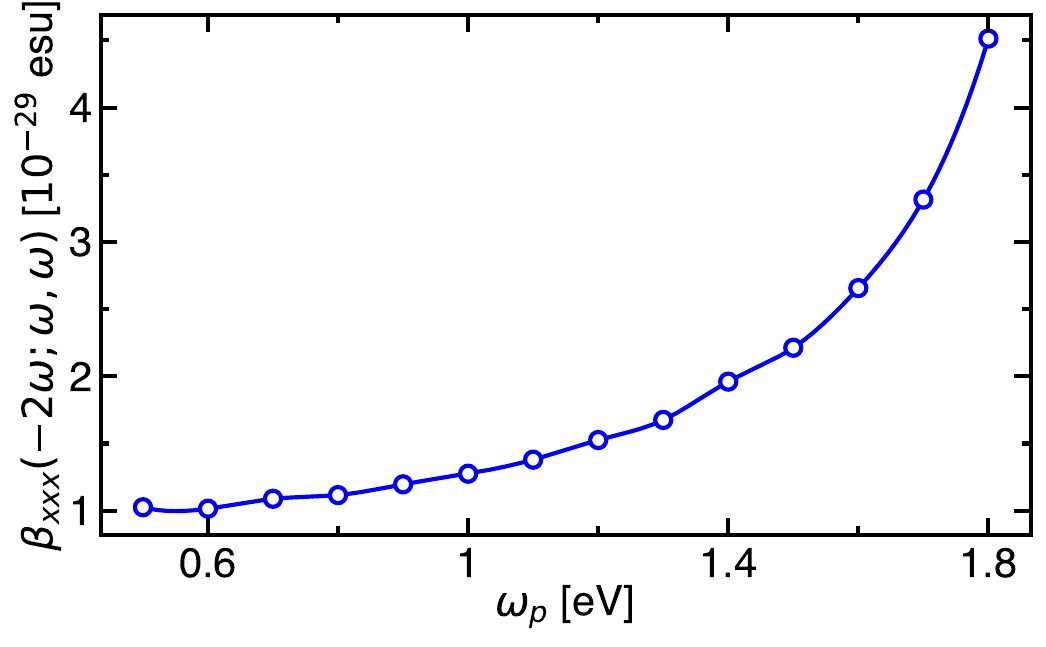}
    \caption{Frequency dependent hyperpolarizability $\beta_{xxx}(-2\omega;\omega,\omega)$ for the para-nitroaniline molecule. 
    Here $\omega_p$ denotes pump frequency.}
   \label{beta_wp_pna_fig}
\end{figure}

\begin{table*}[h]
\centering
\begin{threeparttable}
    \label{mol_pol_table}
    \caption{(Hyper-)Polarizabilities of the para-nitroaniline molecule at $\omega=1.2$ eV computed using RT-TDDFTB. Here, $\alpha(-\omega;\omega)$ is polarizability, $\beta(0;\omega,-\omega)$ is optical refraction, $\beta(-2\omega;\omega,\omega)$ is first hyper-polarizability, $\bar{\gamma}(-\omega;\omega,\omega,-\omega)$ indicates degenerate four-wave mixing, and $\gamma(-3\omega;\omega,\omega,\omega)$ is second hyper-polarizability.
    The values in the last column are taken from \cite{RT-HyperPol-2013-Li}, calculated at $\omega=1.16$ eV using the TD-DFT/6-31G(d) level of theory.
    Here, all results are provided in atomic units (a.u.).
    }
    \setlength{\tabcolsep}{1.3cm}
    \renewcommand{\arraystretch}{1.2}
    \begin{tabular}{|c|c|c|c|}
         \hline
        $\alpha(-\omega;\omega)$ &  $\alpha_{xx}$ & 121 & 143.82\\
                                 &  $\alpha_{yy}$ & 82 & 90.58 \\
                                 &  $\alpha_{zz}$ & 0.2 & 29.54 \\
         \hline
        $\beta(0;\omega,-\omega)$ &  $\beta_{xxx}$ & 1237 & 1755.39\\
                                  &  $\beta_{xyy}$ & -110 & -136.40\\
                                  &  $\beta_{xzz}$ & -0.4 & -3.66\\
                                  &  $\beta_{zxx}$ & 16 & 47.18 \\
                                  &  $\beta_{zzz}$ & 0.02 & 0.95\\
         \hline
        $\beta(-2\omega;\omega,\omega)$ &  $\beta_{xxx}$ & 1766 & 2602.56\\
                                  &  $\beta_{xyy}$ & 175 & 189.24\\
                                  &  $\beta_{xzz}$ & -0.1 & -5.96\\
                                  &  $\beta_{zxx}$ & 22 & 58.19\\
                                  &  $\beta_{zzz}$ & 0.02 & 1.04\\
         \hline
        $\bar{\gamma}(-\omega;\omega,\omega,-\omega)$ &  $\bar{\gamma}_{xxxx}$ & 85606 & 117000.81\\
                                  &  $\bar{\gamma}_{zxxx}$ & 1048 & 1047.77\\
                                  &  $\bar{\gamma}_{xzzz}$ & -0.35 & -27.80\\
                                  &  $\bar{\gamma}_{zzzz}$ & 0.03 & 42.71\\
        \hline
        $\gamma(-3\omega;\omega,\omega,\omega)$ &  $\gamma_{xxxx}$ & 371073 & 428557.48\\
                                  &  $\gamma_{zxxx}$ & 3951 & 4267.92\\
                                  &  $\gamma_{xzzz}$ & -0.9 & -70.85\\
                                  &  $\gamma_{zzzz}$ & 0.03 & 47.17\\
        \hline
    \end{tabular}
  \end{threeparttable}
\end{table*}

\begin{table*}[h]
\centering
\begin{threeparttable}
    \label{mol_pol_table_2}
    \caption{Frequency dependent hyper-polarizability $\beta_{xxx}(-2\omega;\omega,\omega)$ for the para-nitroaniline molecule. 
    Here $\omega_p$ denotes pump frequency.
    }
    \setlength{\tabcolsep}{0.25cm}
    \renewcommand{\arraystretch}{1.2}
    \begin{tabular}{|c|c|c|c|c|c|}
        \hline
         $\omega_p$ [eV] & Expt.$^{\rm a}$ & MP2$^{\rm b}$ & TDDFT/B3LYP$^{\rm c}$ & TDDFT/RT-PBE$^{\rm d}$ & RT-TDDFTB \\
          & & & & & (This Work) \\
         \hline
         \hline
         0.65 & 9.6$\pm$0.5 & & 6.85 & 8.54 & 10.3 \\
         0.905 & 11.8$\pm$0.3 & & 8.15 & 10.57 & 12.0 \\
         1.17 & 16.9$\pm$0.4 & 12.0 & 12.94 & 16.74 & 14.4 \\
         1.364 & 25$\pm$1 & & 17.41 & 22.96 & 17.6 \\
         1.494 & 40$\pm$3 & & 23.09 & 35.37 & 21.7 \\
         \hline
    \end{tabular}
    \begin{tablenotes}
        \item $^{\rm a}$Reference \cite{Hyper-PNA-1983-Garito}
        \item $^{\rm b}$Reference \cite{Hyper-PNA-1993-Rice}
        \item $^{\rm c}$Reference \cite{Hyper-PNA-2002-Agren}     
        \item $^{\rm d}$Reference \cite{Hyper-PNA-2007-Rehr}
    \end{tablenotes}
  \end{threeparttable}
\end{table*}

\clearpage

\bibliographystyle{unsrt}
\bibliography{supp}